\documentclass[twocolumn,twoside,slac_two]{revtex4}
\usepackage{graphicx}
\usepackage{fancyhdr}
\begin{document}

\title{OVRO 40 m Blazar Monitoring Program:\\ Location of the gamma-ray emission region in blazars by the study of correlated variability at radio and gamma-rays}


\author{W. Max-Moerbeck$^{1}$, J. L. Richards$^{2}$, V. Pavlidou$^{3}$, T. J. Pearson$^{1}$, A. C. S. Readhead$^{1}$ on behalf of the \emph{Fermi} LAT collaboration, T. Hovatta$^{1}$, O. G. King$^{1}$ and R. Reeves$^{1}$.}

\affiliation{$^{1}$California Institute of Technology, Pasadena, CA 91125, USA}
\affiliation{$^{2}$Department of Physics, Purdue University, West Lafayette, IN 47907, USA}
\affiliation{$^{3}$Max-Planck-Institut f\"ur Radioastronomie, Auf dem H\"ugel 69, 53121 Bonn, Germany}

\begin{abstract}
Blazars are powerful, variable emitters from radio to gamma-ray wavelengths. Even though the general picture of synchrotron emission at low energies and inverse Compton at the high energy end is well established, many important aspects of these remarkable objects are still not well understood. For example, even the location of the gamma-ray emission region is still not clearly established, with some theories locating it close to the black hole/accretion disk while others place it at parsec scales in the radio jet. Since mid-2007 we have carried out a large scale monitoring program at 15 GHz using the OVRO 40 m telescope. We are currently observing about 1700 blazars twice per week. The sample includes all the \emph{Fermi}-LAT detected blazars north of declination $-20^{\circ}$. Here, we study the existence of correlated variability between these two bands for 86 sources bright enough to be detected weekly by LAT. The existence of correlated variability can be interpreted as an indication of a related spatial locations for the radio and gamma-ray emission, making the evaluation of its statistical significance a key goal of our program. A study of the statistical significance of these cross-correlations is presented along with a discussion of the Monte Carlo simulations used to evaluate them. More information about the conditions on the radio emission zone can be obtained through polarization monitoring which tells us about the configuration of the magnetic fields in this region. To study radio polarization variability we are building KuPol, a radio polarization receiver for the 12 to 18 GHz band that will replace the current total power receiver at the OVRO 40 meter telescope.
\end{abstract}

\maketitle

\thispagestyle{fancy}


\section{INTRODUCTION}

Blazars are a type of active galactic nuclei with a jet oriented close to the line of sight. Their spectral energy distribution is broad and characterized by the presence of two peaks: a low energy one produced by the synchrotron mechanism, and a high energy one produced by inverse Compton scattering of a photon field from the jet, accretion disk, the broad line region or the dust torus. Their emission is variable over a whole electromagnetic spectrum.

Due to these characteristics, a complete understanding of the emission mechanisms in blazars requires their study over a broad range of energies with simultaneous monitoring at different energy bands. This monitoring needs to be independent of the state of the source to avoid biases. Moreover, understanding the statistical properties of different classes requires the study of large samples of objects.

Even though the general picture of synchrotron emission at low energies and inverse Compton at the high energy end is well established, many important aspects of these remarkable objects are still not well understood. Our work concerns one such question: where in the blazar structure is the gamma-ray emission region located? Some theories locate it near the black hole/accretion disk \citep[][]{blandford_and_levinson_1995}, while others place parsecs away down the jet \citep[][]{jorstad_2001}.

The \emph{Fermi} Large Area Telescope (LAT) and other gamma-ray telescopes lack the resolution needed to directly image the gamma-ray emission regions in blazars, but we can investigate their location by studying the connection between the radio and gamma-ray variability. To enable such studies, we are carrying out a  blazar monitoring program with a sample size and cadence matched to \emph{Fermi's} capabilities.

This paper presents a summary of the state of this project as was presented at the time of the symposium. A detailed discussion concerning the data analysis and results will be presented in Max-Moerbeck et al. (in preparation).

\section{OBSERVATIONS}

We are currently monitoring  about 1700 blazars including all the sources north of $-20^{\circ}$ in declination in the Candidate Gamma-ray Blazar Sample \citep[][]{healey+2008} and all \emph{Fermi} detected sources associated with blazars in the first and second LAT AGN catalogs \citep[][]{1LAC_2010, 2LAC_2011} with the same declination limit. The sources are observed in total flux density twice per week at a frequency of 15 GHz with a 3 GHz bandwidth. A typical observation has about 4 mJy of thermal noise with an additional $\sim3\%$ uncertainty coming mostly from pointing errors. A complete description of the program can be found in \citet[][]{richards+2011}.

These observations complement the continuous monitoring of the gamma-ray sky in the band from 100 MeV to 300 GeV by \emph{Fermi}-LAT. In this study, we consider all the gamma-ray sources detected in at least 75\% of the monthly time bins as reported in \citet[][]{2FGL_2012}. This sample contains 86 sources for which we obtain weekly binned gamma-ray light curves for integrated photon flux in the energy range from 100 MeV to 200 GeV covering the time period from 4 August 2008 through 8 August 2011. The radio light curves cover the period from 1 January 2008 through 26 February 2012.

\section{MEASURING THE RADIO/GAMMA-RAY TIME LAGS}

To study the physical significance of cross-correlation peaks we use Monte Carlo simulations that account for the uneven sampling and noise properties of the light curves. The uneven sampling is incorporated using the discrete correlation function \citep[][]{edelson_1988} and the noise properties of the sources are characterized by fitting their power spectral density (PSD) with a simple power law in both bands (P$(f) \propto 1/f^\beta$) using an adaptation of the method presented in \citet[][]{uttley+2002}. Here we present results of a study using populations values of the PSD power law index at both bands. For the gamma-ray band we adopt the average of the results obtained for the mean PSD of the brightest BL Lacs and FSRQs ($\beta_{\gamma} = 1.6$ from \citep[][]{abdo_variability_2010}). For the radio band we use the mean population value we obtain in this work from fits to the radio light curves of the complete OVRO monitored sample ($\beta_{\rm radio} = 2.3$).

The significance of a cross-correlation peak is measured by obtaining an estimate of the distribution of random cross-correlation amplitudes at the given time lag. This distribution is estimated by simulating a large number of independent time series (we use 20,000 in this case) with the appropriate power spectral densities and observational noise.

\section{SIGNIFICANT CORRELATIONS}

\begin{figure*}[t]
\centering
\includegraphics[width=170mm]{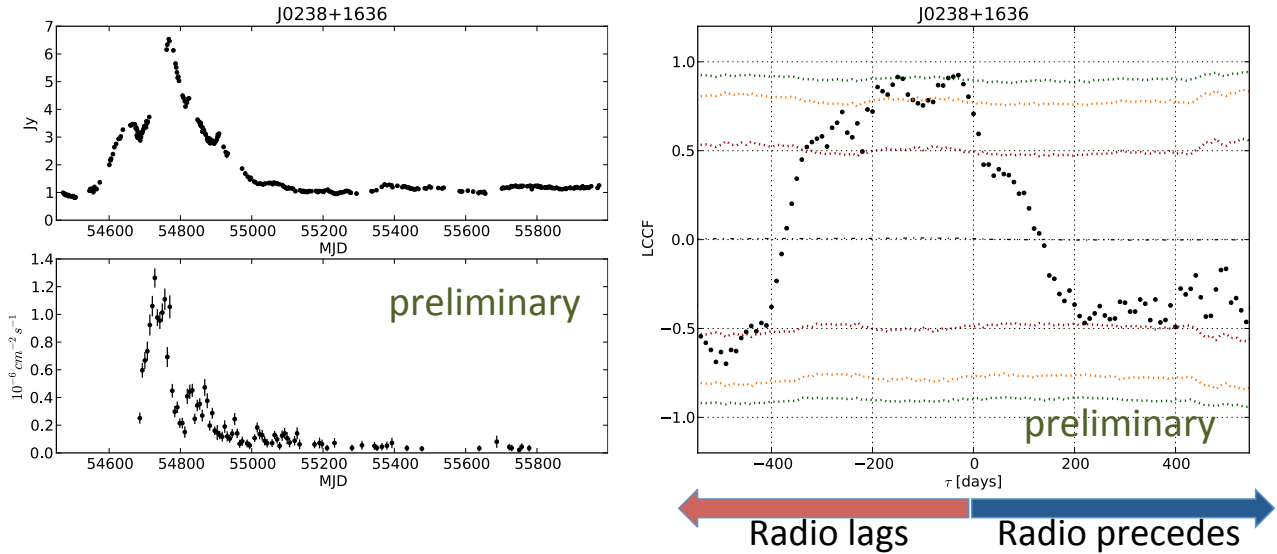}
\caption{Light curves and cross-correlation results for J0238+1636. The most significant peak it at $-30\pm8$ days and has a 99.9\% significance. Color contours on right panel indicate the significance of the cross-correlations with red for $1\sigma$, orange for $2\sigma$ and green for $3\sigma$.} 
\label{plots_J0238+1636}
\end{figure*}

\begin{figure*}[t]
\centering
\includegraphics[width=170mm]{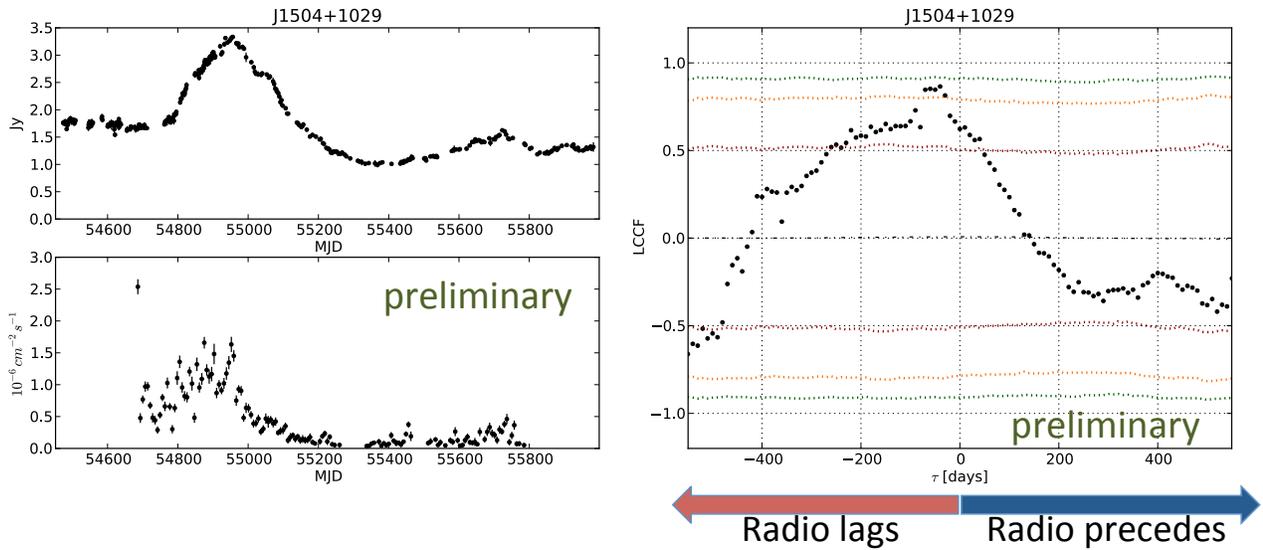}
\caption{Light curves and cross-correlation results for J1504+1029. The most significant peak it at $-40\pm13$ days and has a 98.5\% significance. Color contours as in Figure \ref{plots_J0238+1636}.} 
\label{plots_J1504+1029}
\end{figure*}

Of the 86 sources with good cadence in radio and gamma-ray bands included in this study, only 63 show variability in both bands and are thus included in the cross-correlation analysis. Among these 63 we find one case with $> 3\sigma$ significant correlations, J0238+1636 (Figure \ref{plots_J0238+1636}; 0.2 cases are expected by chance). There are 6 other cases with $> 2\sigma$ significant correlations (3 cases are expected by chance). However, we consider five of these unreliable because of high noise or slow trends. The remaining example, J1504+1029, is shown in Figure \ref{plots_J1504+1029}.

\section{THE CASE OF MRK 421}

\begin{figure*}[t]
\centering
\includegraphics[width=170mm]{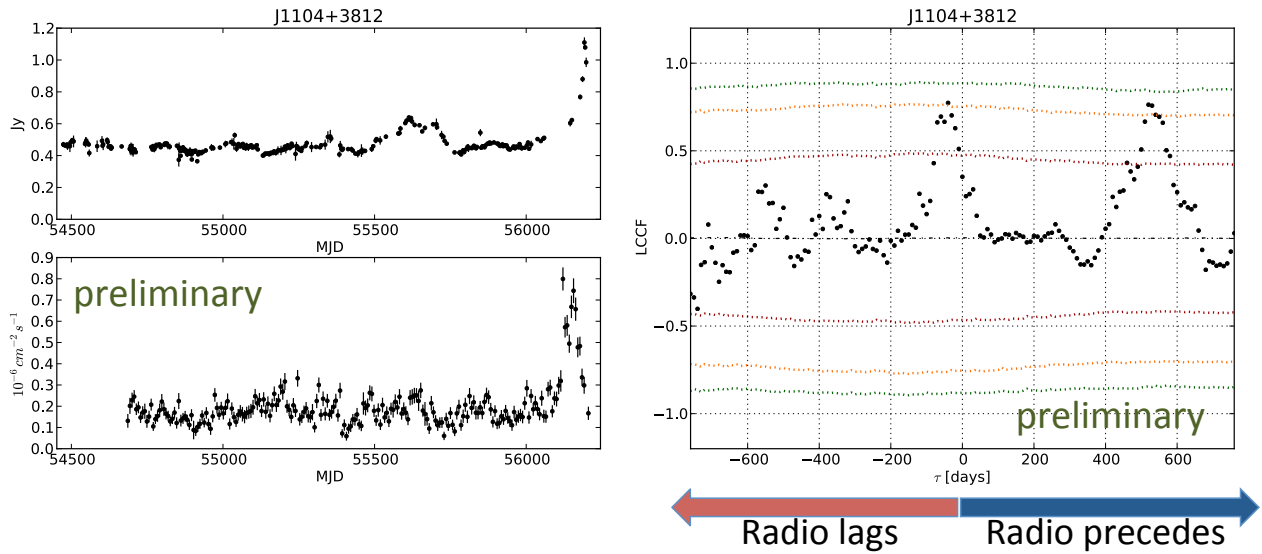}
\caption{Light curves and cross-correlation results for Mrk 421. There are two significant peaks one at $520$ days with 98.8\% significance and a shorter one at $-40$ days with 96.3\% significance.  Color contours as in Figure \ref{plots_J0238+1636}. This lag determination needs to be confirmed with more data including the decaying part of the radio flare.} 
\label{plots_mrk421}
\end{figure*}

The well known blazar Mrk 421 underwent a major gamma-ray flare in July 2012 \citep[][]{fermi_mrk421_atel_2012} which was followed by a major event at 15 GHz, the largest flare since at least 1980 as shown by a comparison to University of Michigan Radio Astronomy Observatory 14.5 GHz data \citep[][]{ovro_mrk421_atel_2012}. When we extend the light curves to include the flaring period we are able to constrain the power spectral density in both bands and find correlated variability with $> 2\sigma$ significance. Preliminary light curves and results of the significance analysis including only the rising part of the radio flare are shown in Figure \ref{plots_mrk421}. This result has to be considered with care as in this case we are extending the period analyzed after observing the flare, thus doing ``a posteriori" statistics.

\section{KUPOL: RADIO POLARIZATION MONITORING}

Blazars produce their radio emission by the synchrotron mechanism which can result in polarized emission. Variations in the polarization can help us understand the evolution of the magnetic fields in the emission regions and its relation to high-energy emission. To study these magnetic fields we are building a new receiver which will also measure linear polarization in the Ku-band. This new receiver, which we call KuPol, will cover the band from 12 GHz to 18 GHz with 16 MHz spectral resolution, which can be used for examining spectral indices and removal of radio frequency interference. This will be an important addition to our program and commissioning is planned for spring 2013.

\section{SUMMARY}

We study the connection between radio and gamma-ray emission in \emph{Fermi} detected blazars and find that significant correlations are present in a small number of these sources with the current data set. The major flare on Mrk 421 shows correlated radio/gamma-ray activity. These cases add support to the idea that radio and gamma-rays are produced in related regions for some sources, but more data are needed to test whether this holds for the entire population. In particular longer radio and gamma-ray light curves will allow us to probe the variability on different time scales, and to incorporate information for multiple flaring periods. This will improve the statistics and the significance of possible correlations.

\bigskip % extra skip inserted
\begin{acknowledgments}
We thank Russ Keeney for his efforts in support of observations at OVRO. The OVRO 40m monitoring program is supported in part by NASA grants NNX08AW31G and NNX11A043G, and NSF grants AST-0808050 and AST-1109911. T.H. was supported by the Jenny and Antti Wihuri foundation. Support from MPIfR for upgrading the OVRO 40 m telescope receiver is also acknowledged.
\end{acknowledgments}

\bigskip % extra skip inserted

\end{document}